*Article*

# Electrophoretic deposition of WS$_2$ flakes on nanoholes arrays – role of the used suspension medium

**Dario Mosconi**[1], **Giorgia Giovannini**[2], **Nicolò Maccaferri**[3], **Michele Serri**[2], **Paolo Vavassori**[4], **Stefano Agnoli**[1] **and Denis Garoli**[2*]

1. Dipartimento di Chimica, Università degli Studi di Padova, Via Marzolo 1, 35131, Padova, Italy
2. Istituto Italiano di Tecnologia – Via Morego, 30, I-16163 Genova, Italy
3. Physics and Materials Science Research Unit, University of Luxembourg, L-1511 Luxembourg, Luxembourg
4. CIC nanoGUNE, Tolosa Hiribidea, 76, E-20018 Donostia-San Sebastian, Spain
* Correspondence: denis.garoli@iit.it;

**Abstract:** Here we optimized the electrophoretic deposition process for the fabrication of WS$_2$–plasmonic nanohole integrated structures. We showed how the conditions used for the site-selective deposition influenced the properties of the deposited flakes. In particular, we investigated the effect of different suspension medium used during the deposition both in the efficiency of the process and in the stability of WS$_2$ –flakes, which were deposited on a ordered arrays of plasmonic nanostructures.

## 1. Introduction

Over the last decade, intensive research efforts have been devoted to the investigation of two-dimensional (2D) materials. Initially dominated by graphene, now the focus is on a broad range of materials like boron nitride (BN), transition metal dichalcogenides (TMDs) such as MoS$_2$ and WS$_2$, transition metal oxides (TMOs) as well as others including black phosphorous, silicene, etc. These materials are now deeply investigated due to their diverse properties and potential applications spanning from photonics, electronics, biosensing and catalysis[1,2]. Among the others, TMDs semiconductors, such as XS$_2$, XSe$_2$, etc, exhibit strong light- matter coupling and possess direct band gaps in the infrared and visible spectral regimes, making them interesting for various applications in optics and optoelectronics[2]. During the last decade researchers has been focused on exploring the potential of the TMDs. However, controlling the optical properties of a XS$_2$ layer(s) is still a great challenge and significant efforts are still required. The active control of XS$_2$ can be, in principle, obtained by utilizing photonic and plasmonic nanostructures.[3–9] Noteworthy, XS$_2$ application in photonics and plasmonics requires to align nanostructures with the 2D material. This can be done depositing nanoparticles or nanostructures over the 2D layer, procedure typically done by means of complex and time-consuming lithographies. An alternative approach consists in the deposition of the 2D layer over the nanostructures, but this requires transfer procedures that are quite challenging and time-consuming.

In this context, we have recently proposed and demonstrated the possibility to control the deposition of single layer MoS$_2$ flakes on metallic nanostructures by means of chemical conjugation and electrophoretic deposition (EPD)[10,11]. The latter, in particular can be applied to any substrates comprising nanoapertures, decorating them with 2D material flakes as either a single or few layers.

EPD is a colloidal process where the suspended charged particles are impelled from the suspension medium to the substrate by an electric field. The process has been extensively applied to 2D materials for several applications, such as sensing, Li-ion battery, anticorrosive coatings, etc[12]. The kinetics of EPD depends on different parameters. Theoretical and modeling studies were carried out to clarify the mechanisms of EPD, including the EPD of 2D materials. Electrochemical parameters such as conductivity, solvents, zeta potential, electric field, concentration, etc. on the EPD of 2D materials were also studied. In fact, EPD is based on the capability of the 2D material flake to acquire an electric charge in the solvent of suspension, here named suspension medium.

In this paper we report on EPD of $WS_2$ flakes on metallic nanoholes. The procedure recently described in our previous work[11], enables site-selective deposition. We observed it's significantly affected by the used suspension medium. For this reason, we investigated different mediums looking at the kinetics of EPD and at the stability of the material in terms of contamination and oxidation. We observed that, with a proper suspension medium it's possible to significantly improve the deposition time meanwhile maintaining the structural composition of the material.

We believe that the method here optimized can be used to prepare structures that can find several interesting applications in the research fields based on 2D materials [4–6,13–19]. For example, here we report three different potential applications such as, fabrication of hybrid nanopores, enhanced Raman spectroscopy and photoluminescence. Numerical simulations are reported to support the observed enhanced light-matter interaction.

**2. Materials and Methods**

**WS$_2$ Exfoliation**

In a glove-box (water < 1 ppm, $O_2$ < 10 ppm), $LiBH_4$ (0.130 g, 6 mmol) and $WS_2$ (0.496 g, 2 mmol) were grounded in a mortar and subsequently transferred in a Schlenk-tube, brought outside of the glove-box and then connected to a Schlenk-line. The sample was heated in a sand bath at 340°C for 5 days under nitrogen. The intercalation product was dropped in 300 ml degassed water in one portion and the suspension was bath-sonicated for 1 h. In order to remove LiOH, the material was washed three times by centrifugation at 10000 rpm (23478g) for 15 min. To select size, the sample was therefore centrifuged at 8000 rpm for 15 min to remove lighter nanosheets and at 1000 rpm for 5 min to settle unexfoliated material.

**X-ray photoemission spectroscopy (XPS)**

XPS is performed in a custom designed ultrahigh vacuum (UHV) system comprising a load lock and an analysis chamber equipped with an Omicron electron analyzer (EA125), and a dual anode X-ray source (omicron DAR 400). The materials to be investigated are drop casted on a Copper support to form a homogeneous thin film. The dried samples are then inserted in the UHV system and left outgassing extensively for a few hours. All XPS measurements are performed under UHV conditions ($<10^{-8}$ mbar) at room temperature using a non monochromated Mg K$\alpha$ source ($h\nu$ = 1253.6 eV) and an electron analyzer pass energy of 20 eV. The binding energy (BE) scale is calibrated using the Au $4f_{7/2}$ core levels (BE = 83.8 eV). XPS multipeak analysis (KolXPD software) was performed using Voigt functions, keeping constant in all the peaks the Full Width Half Maximum (FWHM) and the Gaussian-Lorentzian proportion, which allow to obtain the maximum reliability of the fitting. For the same reason, BEs regarding the same compounds were kept in ± 0.1 eV range from one sample to another.

**Measurement of Dynamics Light Scattering of WS$_2$ Flakes**

DLS experiments were performed using a Malvern Zetasizer nano ZS and the measurements were evaluated using Zetasizer software. Since polar solvents have been used in this study, 1.5 was chosen as the value for f(kα) (Smoluchowski approximation). Thanks to this approximation, the influence of the size of the material in suspension is limited, allowing the use of Henry's equation (therefore DLS) for the evaluation of electrophoretic mobility of nonspherical materials. Data are reported as the average of three measurements (n =3) ± SD. Samples were measured at 25 °C in disposable folded capillary cells (DTS1070) in aqueous dispersants.

**FEM Simulation of $WS_2$ Flakes Integrated over Plasmonic Nanostructures.**

We investigate the plasmonic properties of the structure by means of finite element method (FEM) simulations using an RF module in Comsol Multi- physics taking into account the geometries that were actually fabricated. For the nanohole array, we consider a hole of 100 nm in diameter in a 100 nm thick $Si_3N_4$ membrane with a 50 nm thick Au layer. A top single layer of $WS_2$ has been considered. The refractive indices of Au and $WS_2$ were taken from the works of Rakic et al.[20] and Zhang et al.[21] The model computes the electromagnetic field distribution in each point of the simulation region, enabling the extraction of the quantities plotted below. A single nanostructure was considered by setting the unit cell to be 400 nm wide in both the x- and y- directions, with perfect matching layers (150 nm thick) at the borders. A linearly polarized incident plane wave was assumed to impinge on the structure from the air side.

**Fabrication of Plasmonic Nanoholes**

The fabrication of the metallic nanoholes follows a simple and robust procedure. The substrate was a $Si_3N_4$ membrane (100 nm thick) prepared on a Silicon chip. The 2D holes were prepared by means of FIB milling with a voltage of 30 keV and a current of 80 pA. After milling, a thin layer of gold, ca. 100 nm, was deposited on the top side of the membrane.

**Electrophoretic Deposition**

EPD has been performed following the method illustrate in details by Mosconi et al.[11]. In summary, the substrate is not used as electrode, on the contrary, the nanoholes present in it allow for the through-flow of ions, whilst also representing a barrier to the $WS_2$ flakes, which cannot pass due to their size and are deposited. The nanoholes array prepared on a $Si_3N_4$ membrane is first cleaned in oxygen plasma for 60 s to facilitate the deposition. The sample in the ground state is placed in the microfluidic chamber and the chamber sides are filled with $WS_2$ suspended in different suspension mediums (reported below). A suitable voltage for the required deposition thickness is then applied for few minutes to allow for electrophoretic deposition. The chamber is then opened, and the deposited final sample is rinsed with EtOH.

**3. Results and discussion**

In this work three different suspension mediums have been tested in comparison with water. In particular exfoliated $WS_2$ flakes have been suspended in MES, PBS and B-1. The formulations of these mediums are reported in table 1. MES has been also prepared with different pH with the addition of NaOH (0.1 M). PBS has been tested with different concentrations (0.1; 1 and 10mM). Key aspects to be investigated are the electrophoretic mobility and the effect of the solvents on the structural properties of the $WS_2$.

**Table 1.** Formulation of suspension medium

| Name | Formulation |
|---|---|
| MES | 4-Morpholineethanesulfonic acid (MES) 10 mM |
| PBS | 10mM phosphate buffer. 2.7 mM KCl, 137mM NaCl. |
| B-1 | 1.67 mM Na2SO4, 1 mM NaOH pH 10.8 |

*3.1. Electrophoretic mobility*

One of the parameters that determines the EPD deposition is the net mean charge of the $WS_2$ in the selected suspension medium. For this reason, a detailed analysis of the $WS_2$ flakes electrophoretic mobility (EM) and zeta potential (ζ) is required.

$WS_2$ flakes are characterized by a negative surface charge (-33,976 ± 3,12 mV) when suspended in deionized (DI) water as indicated by the ζ value measured by Dynamic Light Scattering (DLS). ζ is also known as electrokinetic, which is the potential at the surface-fluid interface of a colloid moving under electric field. Once the charged $WS_2$ flakes are dispersed in a suspension medium, ions of the opposite charge will be absorbed at the surface forming a layer of strongly adhered ions (Stern layer) which becomes a diffuse and dynamic layer of a mixture of ions with the increasing distance from the surface. Both Stern and diffuse layer form the electric double layer (EDL) which determines the electrical mobility of the flakes in suspension under electric field and that correspond to the ζ measured by DLS.

Due to the low conductivity of DI water and the consequently low EM for EPD process, the surface charge of $WS_2$ flakes was tested in the different suspension mediums in order to identify the best conditions for the EPD. In particular, the buffers tested have different ionic strength and different pH. Additionally, a 1,67mM sodium sulphate buffer with pH 10.6, reached by addition of NaOH (1mM) have been tested as well. This buffer has been named B-1 and it has tested since it showed good electrophoretic mobility in other different experiments. The values of ζ, EM and Conductivity measured by DLS are presented in Table 2.

**Table 2.** DLS Analyses

| | Z | | EM | | Conductivity | |
|---|---|---|---|---|---|---|
| **MES 10 mM** | Average | SD± | Average | SD± | Average | SD± |
| pH3 | -27.07 | 1.56 | -2.12 | 0.12 | 0.26 | 0.00 |
| pH4 | -31.50 | 0.57 | -2.47 | 0.05 | 0.04 | 0.00 |
| pH5 | -25.53 | 2.93 | -2.00 | 0.23 | 0.12 | 0.00 |
| pH6 | -28.87 | 0.66 | -2.26 | 0.05 | 0.29 | 0.00 |
| pH7 | -28.90 | 0.43 | -2.27 | 0.03 | 0.62 | 0.01 |
| pH8 | -33.07 | 0.62 | -2.59 | 0.05 | 0.79 | 0.01 |
| **PBS pH6.5** | | | | | | |
| 10 mM | -36.87 | 1.37 | -2.89 | 0.11 | 13.33 | 0.33 |
| 1 mM | -34.57 | 1.29 | -2.71 | 0.10 | 1.73 | 0.06 |
| 0.1 mM | -27.40 | 0.62 | -2.15 | 0.05 | 0.16 | 0.00 |
| **Other** | | | | | | |
| B-1 | -37.57 | 0.46 | -2.95 | 0.04 | 0.48 | 0.01 |
| Water | -46.87 | 0.71 | -3.67 | 0.06 | 0.01 | 0.01 |

As noticed from Table 2, the ζ and EM are proportional between each other. DLS software indeed converts the EM measured for a colloidal suspension in the corresponding ζ using Henry equation where ε and η are respectively the dielectric constant and viscosity of the solvent. Henry function, f(kα), instead correlates parameters related to the suspended colloidal *i.e.* the radius of the colloids (α) and to the environment (k).

$$\mu_e = \frac{2\varepsilon z f(k\alpha)}{3\eta} \qquad (1)$$

The variable k, known as Debye parameter, is indeed inversely proportional to the temperature (T) and directly proportional to the ionic strength (I) of the solvent as:

$$\frac{1}{k} = \sqrt{\frac{\epsilon_0 \varepsilon T k_b}{2000 e^2 I N}} \qquad (2)$$

For these considerations, the ζ is strongly related to the ionic strength (IS), pH and therefore conductivity of the solvent-suspension medium. The ionic strength interferes more on the conductivity of the suspension while less differences were observed instead using 10mM MES buffers at different pHs which ranges. In particular, conductivity decreased from 13,433 ± 1,050 mS/cm measured for PBS pH7.4 200mM to 5,517 ± 0,098 mS/cm measured for PBS 5mM. Varying the pH of the solvent and keeping constant the ion's concentration, less difference was observed in terms of conductivity: the lowest and higher value measured are 0,188 ± 0,005 mS/cm and 1,110 ± 0,073 mS/cm which were achieved respectively for the flakes suspension in MES pH5 and MES pH7 respectively.

### 3.2. XPS Analysis

Exfoliated $WS_2$ was analyzed to get the references for metallic 1T phase, which typically shows lower BE values. We observed the typical downshift of about 0.8 eV due to phase change. However, to achieve a satisfying fitting, a fourth component relative to oxidized W was needed: the resulting BE is intermediate between $W_{VI}O_3$ (35.7 eV) and value reported in the literature for $W_{IV}O_2$ (34.0 eV), therefore we can hypothesize the presence of partially oxidized $W_V$ species, formed during the harsh exfoliation process.

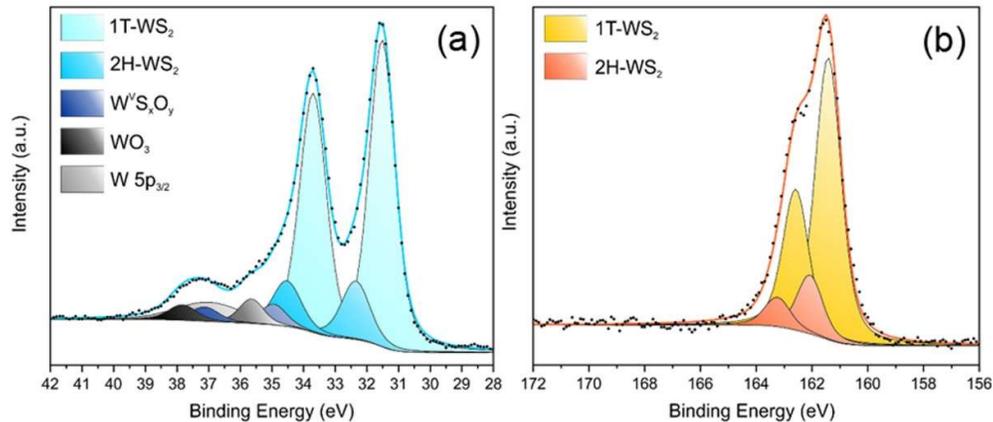

**Figure 1.** XPS spectrum of pristine $WS_2$.

Table 2. XPS of pristine $WS_2$

| W 4f | | | S 2p | | |
|---|---|---|---|---|---|
| Species | BE (eV) | % at. | Species | BE (eV) | % at. |
| $1T\text{-}WS_2$ | 31.5 | 74.2 | $1T\text{-}WS_2$ | 161.4 | 82.8 |
| $2H\text{-}WS_2$ | 32.3 | 15.5 | $2H\text{-}WS_2$ | 162.1 | 17.2 |
| $W^VS_xO_y$ | 34.9 | 5.0 | $SO_x$ | - | - |
| $WO_3$ | 35.7 | 5.3 | | | |

While Table 1 reports the composition obtained from pristine $WS_2$ in $H_2O$, the following Fig. 2 reports the obtained data in the tested suspension mediums.

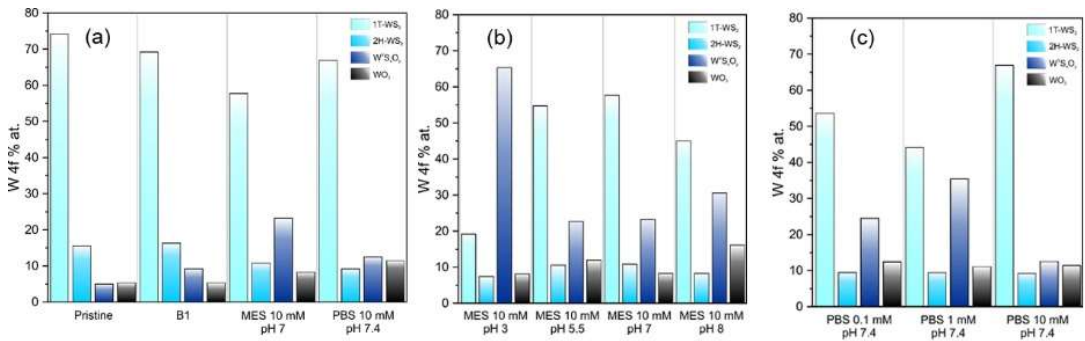

Figure 2. XPS derived composition of $WS_2$ in different suspension mediums.

Comparing the material after the contact with different mediums (Fig. 2a), the solvent B-1 better preserves the properties of the pristine material: the sample shows just a slight partial oxidation of 1T phase to W(V). Among MES samples, the one at neutral pH could maintain the great part of pristine structure, even if a more marked oxidation was observed. The conversion was observed equally in 2H and 1T phases, which were mainly converted to W(V), probably thanks to some interaction with sulfonic acid in MES. PBS 10 mM at pH 7.4, compared to MES, could preserve in a better way the 1T phase, but it is more prone to oxidize to W (IV).

MES can interact chemically with $WS_2$, generating oxysulfides. At acid pH this action is much faster, since about 75% of pristine $1T\text{-}WS_2$ was converted to oxysulfides (Fig. 2b). The effect is then attenuated increasing the pH, in particular at neutral, where the material is substantially preserved. Going to slightly basic pH values, the material got more damaged and the oxidation to $WO_3$ doubled with respect to sample treated at pH 7.

*3.3. Site selctive EPD*

Figure 3 reports the result of EPD performed on WS2 flakes dispersed in H2O. The process has been done applying a voltage of 25V for 5 minutes. As can be seen, the array of metallic holes has been decorated with low uniformity. This result confirms our previous findings achieved with $MoS_2$[11].

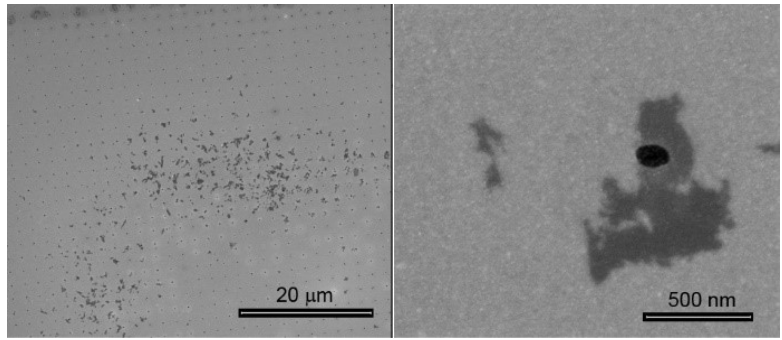

**Figure 3.** EPD of WS$_2$ suspended in H$_2$O.

DLS data suggested that ζ should be considered along with the conductivity of the sample in defining the best conditions for EPD with high kinetics. As expected indeed high voltage were required in order to achieve EDP of flakes dispersed in DI water (25V). In such environment, the conductivity was indeed very low due to the absence of ions (0,021 ± 0,005 mS/cm) even though the ζ was the highest measured for WS$_2$ flakes (-33,967 ± 3,120 mV).

On the contrary, the low EPD kinetics observed in H$_2$O can be overcome by using other suspension mediums. The following figures illustrate the results obtained with B1, MES and PBS by applying a voltage of 3 V for 5 minutes. It is important to note that with this low potential it was not possible to achieve an EPD in H$_2$O.

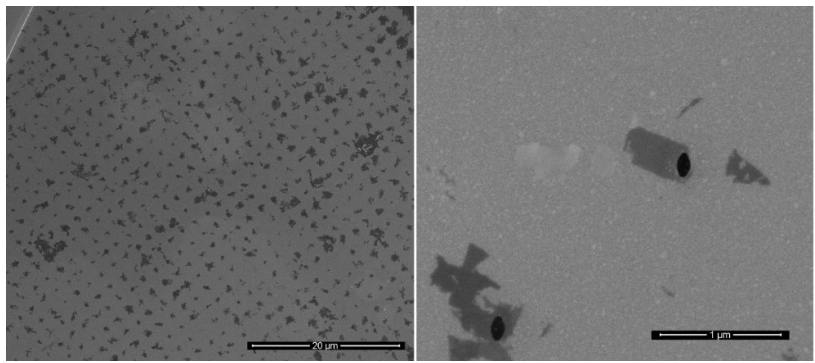

**Figure 4.** EPD of WS$_2$ suspended in B-1.

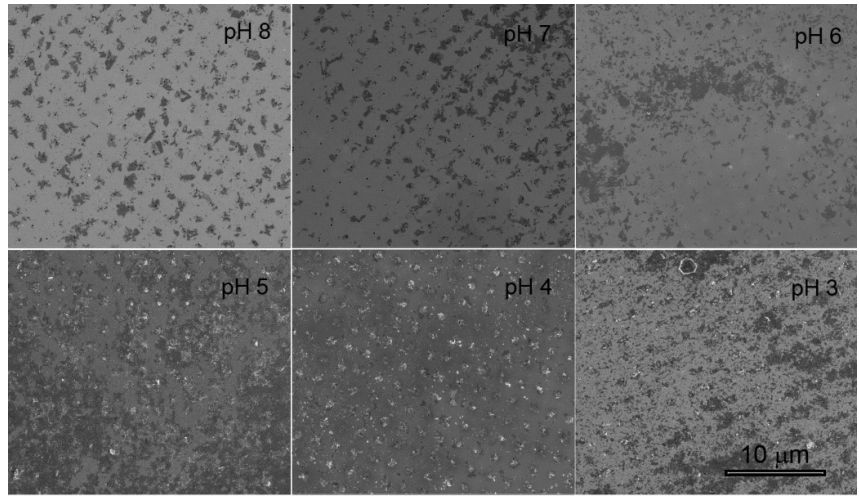

**Figure 5.** EPD of WS$_2$ suspended in MEF with different pH.

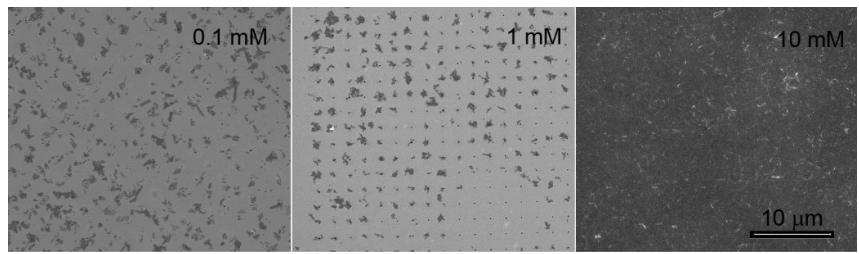

**Figure 6.** EPD of WS$_2$ suspended in PBS at different concentrations.

As can be seen from Fig. 4, 5 and 6 the kinetics of EPD is strongly related with the properties of the medium used. In particular, results showed that B-1 enables a reproducible and ordered deposition of thin flakes over the holes array. Considering the observed good stability of the WS$_2$ in this solvent in terms of oxidation we considered this solvent a better alternative to H$_2$O for EPD process. MES and PBS lead to an efficient EPD on the holes array if used at the suitable pH and concentration. In particular, decreasing the pH of MES we obtained an increasing kinetics of EPD. However, considering the oxidation induced by this solvent (Fig. 2b) it can be effective in WS$_2$ deposition only with a pH of 7. Finally, in the case of PBS we observed the lower reproducibility in the EPD with a significant increase of deposited material once a concentration of 10mM was used. Considering the high oxidation induced by this solvent, it should not be considered for WS$_2$ EPD.

*3.4. Examples of applications*

As mentioned, TMDs and WS$_2$ in particular, find significant application in several fields, from catalysis, to photonics and optoelectronics. Our EPD protocol can be applied to prepare integrated (hybrid) plasmonic nanostructures enabling light-matter interaction studies, such for instance enhanced Raman (SERS) and Photoluminscence (PL). Moreover, integrated 2D material - plasmonic nanostructures can find application in nanopore technology for single molecule detection[22] and an easy method for nanopore fabrication in 2D layers has been recently proposed by us[10]. In these regards, here we report three examples where our EPD of WS$_2$ in B-1 suspension medium has been used for enhanced spectroscopies. In particular a sub 10 nm pore has been prepared by means of Focused Ion Beam (FIB) serial milling (Fig. 7a). The array of nanoholes decorated with WS2 has been mapped by means of a Raman spectrometer (Fig. 7b) and finally, the PL of the array has been recorded after a thermal annealing of the sample in order to induce the 1T-2H transition phase on the flakes.

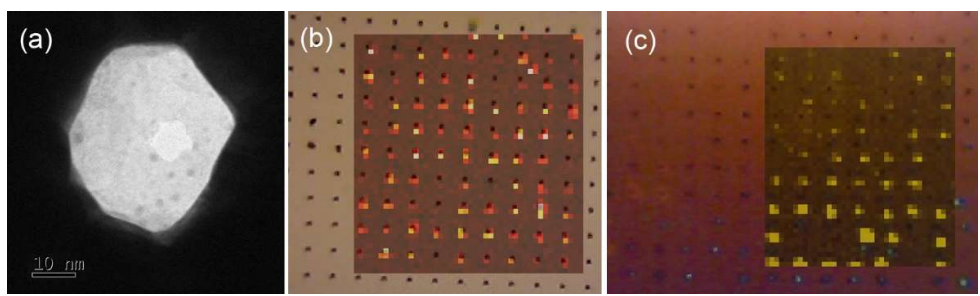

**Figure 7.** (a) Nanopore prepared into the 2D layer; (b) Raman Map, integrated on the $A_{1g}$ mode at 410 cm$^{-1}$; (c). PL map – excitation at 532 nm – emission integrated between 600 and 700 nm.

*3.4. Numerical simulations – Enhanced electromagnetic field confinement*

Computer simulations (see Methods for details) can be used to better evaluate potential enhancement effects. In particular, we considered a gold nanohole covered with a single layer WS$_2$ flake. Into the 2D layer we designed a pore with a diameter of 5 nm. Figure 8 illustrates the obtained field intensity confinement / enhancement that can be achieved. As noticible, the electromagnetic field intensity is highly confined inside the nanohole thanks to the high refractive index of the WS$_2$. This, as illustrated in our Raman and Photoluminesce experiments, enables enhanced light-matter interaction.

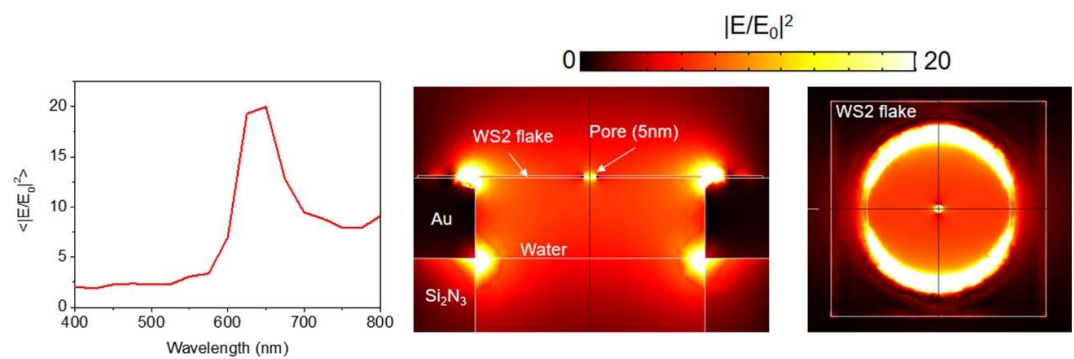

**Figure 8.** Finite Element Method (FEM) simulations of the investigated structures covered with one mono-layer of WS$_2$ (simulations performed at excitation wavelength 633 nm).

**4. Conclusions**

In the presented work we proved that taking under consideration the properties of both colloidal material and solvent it is possible to select parameters by which an efficient and controlled EDP on nano-holes can be easily achieved. Moreover, we proved that the presence of salts in the buffers can interfere with the physical properties of the material. The obtained integrated nanostructures can find interesting applications in enhanced light-matter interaction such as PL and Raman. Finally, our fabrication method can be applied to prepare plasmonic nanopores for potential single molecule studies.

**Author Contributions:** D.M. performed the exfoliation and the XPS analyses; G.G. performed the DLS analyses; N.M. performed the numerical analyses; M.S. supported in EPD; P.V., S.A. and D.G. supervised the work.

**Conflicts of Interest:** The authors declare no conflict of interest.

10 of 11## References

1. Stanford, M.G.; Rack, P.D.; Jariwala, D. Emerging nanofabrication and quantum confinement techniques for 2D materials beyond graphene. *npj 2D Mater. Appl.* **2018**, *2*.
2. Mueller, T.; Malic, E. Exciton physics and device application of two-dimensional transition metal dichalcogenide semiconductors. *npj 2D Mater. Appl.* **2018**, *2*, 1–12.
3. Sun, B.; Wang, Z.; Liu, Z.; Tan, X.; Liu, X.; Shi, T.; Zhou, J.; Liao, G. Tailoring of Silver Nanocubes with Optimized Localized Surface Plasmon in a Gap Mode for a Flexible MoS 2 Photodetector. *Adv. Funct. Mater.* **2019**, *1900541*, 1–8.
4. Ma, C.; Yan, J.; Huang, Y.; Yang, G. Photoluminescence manipulation of WS2 flakes by an individual Si nanoparticle. *Mater. Horizons* **2019**, *6*, 97–106.
5. Gong, L.; Zhang, Q.; Wang, L.; Wu, J.; Han, C.; Lei, B.; Chen, W.; Eda, G.; Goh, K.E.J.; Sow, C.H. Emergence of photoluminescence on bulk MoS2 by laser thinning and gold particle decoration. *Nano Res.* **2018**, *11*, 4574–4586.
6. Shi, J.; Liang, W.Y.; Raja, S.S.; Sang, Y.; Zhang, X.Q.; Chen, C.A.; Wang, Y.; Yang, X.; Lee, Y.H.; Ahn, H.; et al. Plasmonic Enhancement and Manipulation of Optical Nonlinearity in Monolayer Tungsten Disulfide. *Laser Photonics Rev.* **2018**, *12*, 1–7.
7. Kleemann, M.E.; Chikkaraddy, R.; Alexeev, E.M.; Kos, D.; Carnegie, C.; Deacon, W.; De Pury, A.C.; Große, C.; De Nijs, B.; Mertens, J.; et al. Strong-coupling of WSe2 in ultra-compact plasmonic nanocavities at room temperature. *Nat. Commun.* **2017**, *8*.
8. Su, P.-H.; Shih, C.-K.; Tsai, Y.; Johnson, A.D.; Ekerdt, J.G.; Hu, S.; Cheng, F. Enhanced Photoluminescence of Monolayer WS 2 on Ag Films and Nanowire–WS 2 –Film Composites . *ACS Photonics* **2017**, *4*, 1421–1430.
9. Kang, Y.; Najmaei, S.; Liu, Z.; Bao, Y.; Wang, Y.; Zhu, X.; Halas, N.J.; Nordlander, P.; Ajayan, P.M.; Lou, J.; et al. Plasmonic Hot Electron Induced Structural Phase Transition in a MoS 2 Monolayer. *Adv. Mater.* **2014**, *26*, 6467–6471.
10. Garoli, D.; Mosconi, D.; Miele, E.; Maccaferri, N.; Ardini, M.; Giovannini, G.; Dipalo, M.; Agnoli, S.; De Angelis, F. Hybrid plasmonic nanostructures based on controlled integration of MoS 2 flakes on metallic nanoholes. *Nanoscale* **2018**, *10*, 17105–17111.
11. Mosconi, D.; Giovannini, G.; Jacassi, A.; Ponzellini, P.; Maccaferri, N.; Vavassori, P.; Serri, M.; Dipalo, M.; Darvill, D.; De Angelis, F.; et al. Site-Selective Integration of MoS2 Flakes on Nanopores by Means of Electrophoretic Deposition. *ACS Omega* **2019**, *4*, 9294–9300.
12. Ma, Y.; Han, J.; Wang, M.; Chen, X.; Jia, S. Electrophoretic deposition of graphene-based materials: A review of materials and their applications. *J. Mater.* **2018**, *4*, 108–120.
13. Chen, X.; Wang, H.; Xu, N.S.; Chen, H.; Deng, S. Resonance coupling in hybrid gold nanohole–monolayer WS2 nanostructures. *Appl. Mater. Today* **2019**, *15*, 145–152.
14. Hu, H.; Yang, X.; Guo, X.; Khaliji, K.; Biswas, S.R.; García de Abajo, F.J.; Low, T.; Sun, Z.; Dai, Q. Gas identification with graphene plasmons. *Nat. Commun.* **2019**, *10*, 1–7.
15. Lu, Q.; Yu, L.; Liu, Y.; Zhang, J.; Han, G.; Hao, Y. Low-Noise Mid-Infrared Photodetection in BP / h-BN / Graphene van der Waals Heterojunctions. **2019**, 1–8.
16. Zhang, Y.; Kartashov, Y. V.; Zhang, Y.; Torner, L.; Skryabin, D. V. Inhibition of tunneling and edge state control in polariton topological insulators. *APL Photonics* **2018**, *3*.
17. Song, H.; Liu, J.; Liu, B.; Wu, J.; Cheng, H.M.; Kang, F. Two-Dimensional Materials for Thermal Management Applications. *Joule* **2018**, *2*, 442–463.

1.